# Influence of chemical structure on the stability and the conductance of porphyrin single-molecule junctions

*Mickael L. Perrin\*, Ferry Prins\*, Christian A. Martin\*, Ahson J. Shaikh†, Rienk Eelkema†, Jan H. van Esch†, Tomas Briza‡, Rober Kaplanek‡, Vladimir Kral‡, Jan M. van Ruitenbeek§, Herre S. J. van der Zant\*, Diana Dulić\**

\* Kavli Institute of Nanoscience, Delft University of Technology, Lorentzweg 1, 2628 CJ Delft, The Netherlands

† Department of Chemical Engineering, Delft University of Technology, Julianalaan 136, 2628 BL Delft, The Netherlands

‡ Institute of Chemical Technology, Faculty of Chemical Engineering, Technická 5, 166 28 Prague 6, Czech Republic

§ Kamerlingh Onnes Laboratory, Leiden University, Niels Bohrweg 2, 2333 CA Leiden, The Netherlands

**Porphyrin molecules can form stable single molecule junctions without anchoring groups. Adding thiol end groups and pyridine axial groups yields more stable junctions with an increased spread in low-bias conductance. This is a result of different bridging geometries during breaking, the stability of which is demonstrated in time-dependent conductance measurements. This is in strong contrast with rod like molecules which show one preferential binding geometry.**

The idea of using porphyrin molecules as building blocks of functional molecular devices has been widely investigated [1,2]. The structural flexibility and well-developed synthetic chemistry of porphyrins allows their physical and chemical properties to be tailored by choosing from a wide library of macrocycle substituents and central metal atoms. Nature itself offers magnificent examples of processes that utilize porphyrin derivatives, such as the activation and transport of molecular oxygen in mammals and the harvesting of sunlight in plant photosynthetic systems. In order to exploit the highly desirable functionality of porphyrins in artificial molecular devices, it is imperative to understand and control the interactions that occur at the molecule-substrate interface. Such interactions largely depend on the electronic and conformational structures of the adsorbed molecules, which can be studied using techniques such as scanning tunneling microscopy (STM) [3-7], UV Photoemisson Spectroscopy (UPS) [8], X-Ray Photoemission Spectroscopy (XPS) [2] and on a theoretical level with density functional theory (DFT) [9]. Recent studies on conjugated rod-like molecules have shown that molecular conductance measurements can be significantly affected by the binding geometry [10], coupling of the π-orbitals to the leads [11] or π-π stacking between adjacent molecules [12]. In this Communication, we study the interaction of laterally extended π-conjugated porphyrin molecules with the electrodes by means of time and stretching-dependent conductance measurements on molecular junctions. We further investigate strategies to reduce interactions of the molecular π-electrons with the metal electrodes by modifying the chemical structure of the porphyrin molecules.

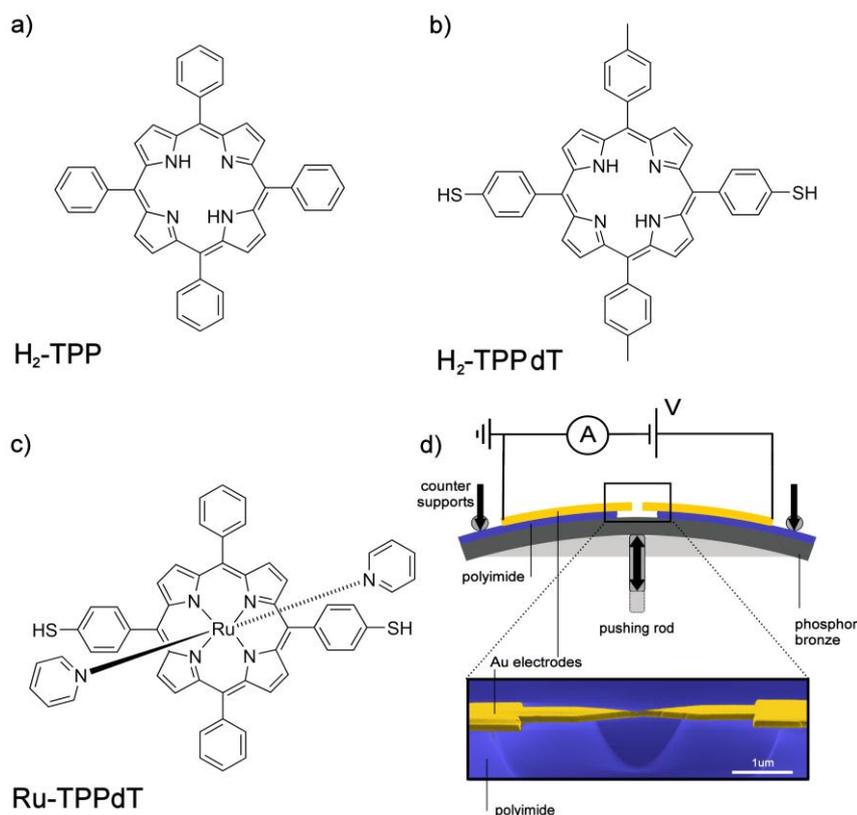

*Figure 1.* (a) $H_2$-TPP (b) $H_2$-TPPdT (c) Ru-TPPdT (d) Top: lay-out of the mechanically controllable break junction (MCBJ). Bottom: scanning electron micrograph of a MCBJ device (colorized for clarity). The scale bar shows that the suspended part of the electrode bridge is about 1.5 μm long.

We have used the series of molecules represented in Fig. 1a-c to examine the influence of the molecular structure on the formation of porphyrin single-molecule junctions. Since the thiol group is most commonly used to contact rod-like molecules to form straight molecular bridges [13], we first compared the porphyrin molecule (5,10,15,20-tetra(p-phenyl)porphyrin, abbreviated as $H_2$-TPP) without thiol termination (Fig. 1a) to a nearly identical molecule with two thiol groups on opposite sides of the molecule (5,15-di(p-thiophenyl)-10,20-di(p-tolyl)porphyrin, abbreviated as $H_2$-TPPdT, Fig. 1b). To investigate the influence of the molecular backbone geometry on the junction formation we further studied a thiol terminated porphyrin molecule with two bulky pyridine axial groups attached via an octahedral $Ru^{II}$-ion ($[Ru^{II}(5,10,15,20$-tetra(p-phenyl)porphyrin)$(pyr)_2]$ abbreviated as Ru-TPPdT, Fig. 1c). Due to steric hindrance the pyridine groups reduce the direct interaction of the metal electrodes with the π-face of the porphyrin. A similar strategy has used previously [4].

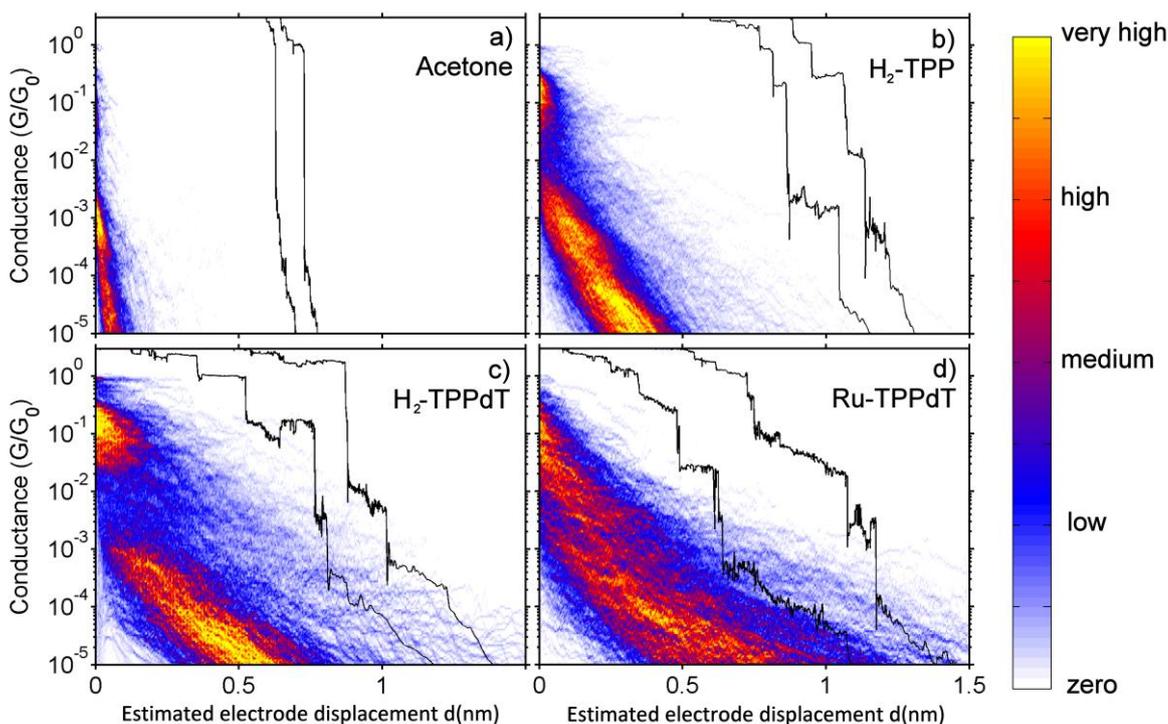

*Figure 2*. Trace histograms constructed from 1000 breaking traces for junctions exposed to (a) pure acetone (b) $H_2$-TPP (c) $H_2$-TPPdT and (d) Ru-TPPdT. Regions of high counts represent the most probable breaking behavior of the contact. The black curves are examples of individual breaking traces (offset for clarity). For the construction of the histograms d=0 for each curve was set to the point where the conductance drops sharply below 1 $G_0$. All histograms were taken at a bending speed of 30μm/s (on the scale of the electrodes, this bending is translated to a stretching on the order of 1.8 nm/s) and a bias voltage of 150 mV. No data selection schemes have been utilized. For additional trace histograms see Supporting Information. For (a) we found a decay constant of 2 Å$^{-1}$.

Prior to electrical characterization, molecules were deposited using self-assembly from solution. To study the conductance of these molecules we used lithographic mechanically controllable break junctions (MCBJ) in vacuum at room temperature. The lay-out of a MCBJ device in a three-point bending mechanism is shown in Fig. 1d. Details concerning the synthesis of the molecule and the experimental procedures are given in the Supporting Information. Sets of 1000 consecutive breaking traces from individual junctions were analyzed numerically to construct 'trace histograms' of the conductance ($\log_{10}(G)$ versus the electrode displacement d) [14, 15]. This statistical method maps the breaking dynamics of the junctions beyond the point of rupture of the last monatomic gold contact (defined as d=0), which has a conductance of one quantum unit $G_0=2e^2/h$. Areas of high counts represent the most typical breaking behavior of the molecular junctions. Fig. 2 presents 'trace histograms' as well as examples of individual breaking traces for an acetone reference sample (a) and junctions exposed to $H_2$-TPP (b), $H_2$-TPPdT (c) and Ru-TPPdT (d). For all three porphyrin molecules as well as for the reference sample which was exposed to pure acetone several junctions have been measured (see Supporting Information). Here, we only show a typical set of junctions.

In the junction which was exposed to the pure solvent without porphyrin molecules (Fig. 2a), the Au-bridge initially gets stretched until a plateau around the conductance quantum (G ~ $G_0$) is observed (only visible in the individual traces shown in black), which corresponds to a monatomic contact [14]. Upon further stretching, the gold-gold contact is broken and the conductance decreases abruptly to ~$10^{-3}$ $G_0$. Beyond this point, electron tunneling between the electrodes leads to a fast conductance decay with stretching (visible as the orange tail), as expected for tunnelling across a vacuum barrier.

In contrast, introducing the porphyrin molecules by self-assembly on the junctions leads to pronounced plateaus at different conductance values in the sub-$G_0$ regime. These plateaus can be flat or sloped[10,16]. The representative breaking traces that are included in Fig. 2b-d display a set of such plateaus. Averaging over 1000 traces does not lead to a narrow region of high counts in the histograms, in contrast to measurements on rod-like molecules [14, 15,16,17]. In the trace histogram of $H_2$-TPP (Fig. 2b), however, there are two distinct regions with high counts; a high-conductance region (HCR) around $10^{-1}$ $G_0$, and a sloped low-conductance region (LCR) below $10^{-3}$ $G_0$. For $H_2$-TPPdT (Fig. 2c) both the HCR and LCR are longer than for $H_2$-TPP; thus, adding the thiol end groups to $H_2$-TPP increases the plateau length. Adding the thiol to $H_2$-TPP also reduces the slope of the LCR. In Ru-TPPdT (Fig. 2d) the HCR and LCR cannot be distinguished anymore: only one long region sloping from $10^{-1}$ $G_0$ to $10^{-5}$ $G_0$ is present, which has a shallower slope compared to $H_2$-TPP and $H_2$-TPPdT, and an increased length.

Additional information about molecular junction configurations can be gained by measuring the evolution of the junction conductance over large time intervals at fixed electrode distances [18]. To obtain such 'time traces', we opened the junction in small steps using a servo motor at 77K. This low temperature enhances the stability due to a reduced surface diffusion without causing extensive changes in molecular conductance [19]. We then measured the conductance in the range from 1 $G_0$ to $10^{-4}$ $G_0$ at various fixed electrode spacings and for time periods exceeding several hours. Due to the exceptional control over the electrode separation in the MCBJ [18,19], conductance jumps in these measurements can be attributed to reconfigurations of the molecule in the junction rather than mechanical interference of the setup.

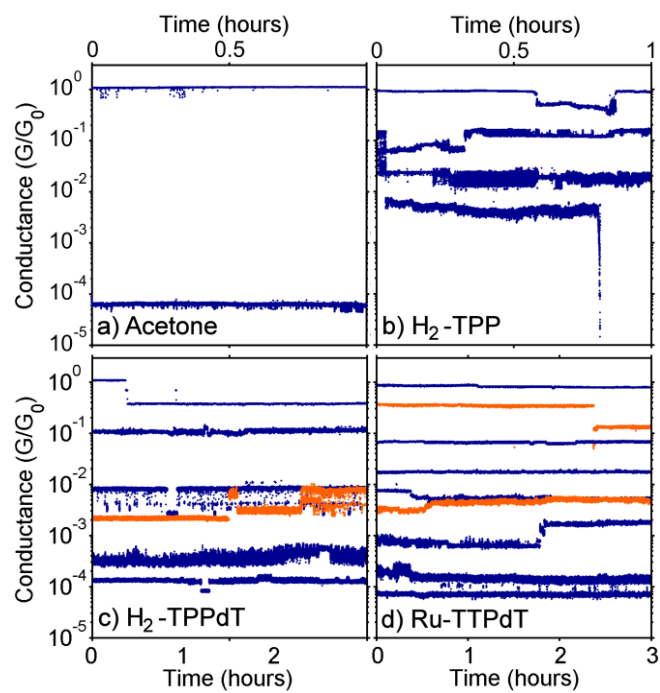

*Figure 3.* Conductance versus time plots for (a) acetone (b) $H_2$-TPP (c) $H_2$-TPPdT and (d) Ru-TPPdT taken at bias voltage of 150 mV. The traces originate from the same opening event and are taken at an interval of 50 pm. Several lines are displayed in orange for clarity. The observed conductance variations (noise on the trace) are due to small changes in the molecular configuration. Larger jumps are attributed to spontaneous reconfiguration of the molecular junction.

Typical time traces of junctions exposed to pure acetone, $H_2$-TPP, $H_2$-TPPdT and Ru-TPPdT are presented in Fig. 3a-d. As can be seen in Fig. 3a, a stable contact is formed at 1 $G_0$ in the clean junction. A slight increase in the electrode spacing leads to abrupt breaking of the contact to a conductance value around $10^{-4}$ $G_0$. In contrast to this sharp drop, the presence of porphyrin molecules leads to the formation of junctions with conductances in the region between 1 $G_0$ and $10^{-3}$ $G_0$ ($H_2$-TPP Fig. 3b, $H_2$-TPPdT Fig. 3c and RuTPPdT Fig. 3d). Below $10^{-3}$ $G_0$, the $H_2$-TPP molecule (Fig. 3b) exhibits a sudden conductance drop to below $10^{-5}$ $G_0$. In contrast, thiol terminated $H_2$-TPPdT and Ru-TPPdT (Fig. 3c-d) form molecular junctions over the whole range between 1 $G_0$ and $10^{-4}$ $G_0$. The observations support the conclusions drawn from the trace histograms: adding thiol and pyridine groups increases the junction stability and leads to a variety of molecular geometries with different conductance values. To examine the contribution of π-stacking of multiple molecules, we measured trace histograms on junctions exposed to meso-tetra-(3,5-di-t-butylphenyl)porphine abbreviated as TBP. Due to steric hindrance of the butyl groups, the possibility of pi-stacking between two molecules is reduced. Resulting trace histograms are shown in the Supporting Information. There is no qualitative difference between the trace histogram of TBP and TPP molecules, which demonstrates insignificant influence of π-stacking to the formation of different configurations.

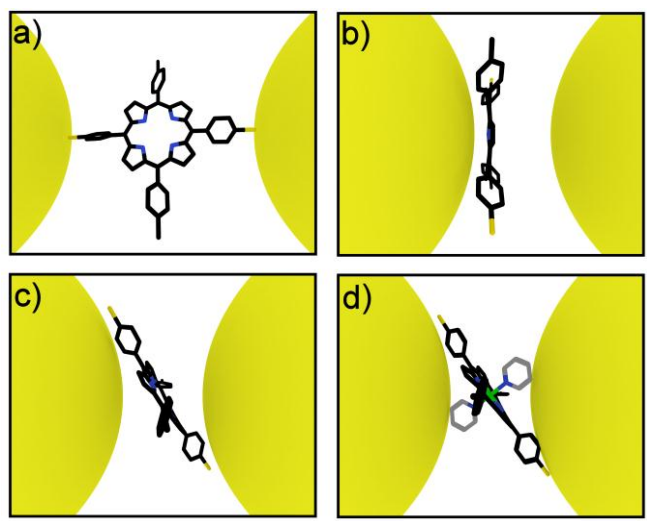

*Figure 4.* Illustration of four possible configurations of a porphyrin molecule in the junction. (a) Bridging configuration of $H_2$-TPPdT (b) STM-like configuration of $H_2$-TPPdT (c) Intermediate configuration of $H_2$-TPPdT (d) Intermediate configuration of Ru-TPPdT stabilized by two bulky pyridine groups (shown in grey).

A closer inspection of the time traces reveals interesting information on the behavior of the molecule in the junction. Since the electronic noise is much smaller than the width of the observed bands of the measured conductance values, we conclude that what appears to be noise in the time traces is in fact a sign of small variations in the molecular configuration [18]. Large jumps in the conductance (Fig. 3c-d) indicate that a molecular junction can spontaneously change configuration. In Fig. 3c random telegraph noise between two conductance values is observed indicating the possibility of forming several meta-stable configurations. Furthermore, Fig. 3d shows that a particular junction conductance can be reached from different starting values (blue and orange traces in the middle of the figure).

The presence of a large range of molecular adsorption geometries contrasts with most studies on long rod-like molecules, which typically assume a straight bridging configuration of the molecule with both thiol groups connected to the electrodes (see Fig. 4a). For $H_2$-TPPdT, such a configuration could be expected. However, the high affinity of the porphyrin π-cloud for metal surfaces likely stabilizes other configurations, such as sketched in Fig. 4b and 4c. A stable junction configuration can be formed without thiol groups [20] and recent break junction experiments have demonstrated that benzene moieties can bind directly to gold electrodes [21,22]. Furthermore, it has been reported that the lateral coupling of π orbitals to the electrodes influences the single-molecule conductance of rod-like molecular wires [23,11]. Such coupling becomes more likely when laterally extended molecules are probed. In addition, tetraphenyl porphyrin molecules have internal degrees of freedom, which can influence the charge transport on a single molecule level [24]. STM studies have indicated that tetraphenyl porphyrins can bind to Au(111) through an interaction with their phenyl side groups [6]. On gold surfaces their conformation can change through rotations of side groups and buckling of the center [5]. Such variations in the adsorption geometry, which can be expected to lead to different conductance values, are likely the origin of the variety of junction configurations that we observe.

In summary, we have electrically probed different junction configurations of porphyrin molecules. We have demonstrated that porphyrin molecules can form stable bridging molecular junctions even without thiol anchoring groups. Adding the thiol end groups and pyridine axial groups to the porphyrin backbone, respectively, increases the stability of the junctions and leads to an increased spread in conductance. This is a result of the formation of different junction configurations. To enable the reliable formation of stable porphyrin junctions, molecules with well-defined adsorption geometries will be required. Quantum chemistry calculations could yield more insight into their design as well as the configurations and the conductance of porphyrin single-molecule junctions. Finally, we expect that multiple junction configurations with considerable variation in the measured conductance values can also be observed for other non-rod-like molecules.

**Acknowledgements**

This research was carried out with financial support from the Dutch Foundation for Fundamental Research on Matter (FOM) and the VICI (680-47-305) grant from The Netherlands Organisation for Scientific Research (NWO). The authors would like to thank Prof. dr. Robert Metzger and Dr. Graeme R. Blake for careful reading of the manuscript.